\documentclass[apj]{emulateapj}
\usepackage{xcolor}
\usepackage{amsmath}
\usepackage{amsfonts}
\usepackage{amssymb}
\usepackage{natbib}

\begin{document}

\title{Misalignment of Outflow Axes in the Proto-Multiple Systems in Perseus}
\author{Katherine I.\  Lee\altaffilmark{1},
Michael M.\ Dunham\altaffilmark{1},
Philip C.\ Myers\altaffilmark{1},
H\'{e}ctor G.\ Arce\altaffilmark{2},
Tyler L.\ Bourke\altaffilmark{3,1},
Alyssa A.\ Goodman\altaffilmark{1},
Jes K.\ J\o rgensen\altaffilmark{4},
Lars E.\ Kristensen\altaffilmark{1},
Stella S.\ R.\ Offner\altaffilmark{5},
Jaime E.\ Pineda\altaffilmark{6},
John J.\ Tobin\altaffilmark{7},
and
Eduard I.\ Vorobyov\altaffilmark{8,9}
}
\affil{$^{1}$Harvard-Smithsonian Center for Astrophysics, Cambridge, MA 02138, USA; katherine.lee@cfa.harvard.edu}
\affil{$^{2}$Department of Astronomy, Yale University, New Haven, CT 06520, USA}
\affil{$^{3}$SKA Organization, Jodrell Bank Observatory, Lower Withington, Macclesfield, Cheshire SK11 9DL, UK}
\affil{$^{4}$Niels Bohr Institute and Center for Star and Planet Formation, Copenhagen University, DK-1350 Copenhagen K., Denmark}
\affil{$^{5}$Department of Astronomy, University of Massachusetts, Amherst, MA 01003, USA}
\affil{$^{6}$Max-Planck-Institut f\"{u}r extraterrestrische Physik, 85748 Garching, Germany}
\affil{$^{7}$Leiden Observatory, Leiden University, Leiden, The Netherlands}
\affil{$^{8}$Department of Astrophysics, The University of Vienna, Vienna, A-1180, Austria}
\affil{$^{9}$Research Institute of Physics, Southern Federal University, Rostov-on-Don, 344090, Russia}

\begin{abstract}
We investigate the alignment between outflow axes in nine of the youngest binary/multiple systems in the Perseus Molecular Cloud.
These systems have typical member spacing larger than 1000 AU.
For outflow identification, we use $^{12}$CO(2-1) and $^{12}$CO(3-2) data from a large survey with the Submillimeter Array: Mass Assembly of Stellar Systems and their Evolution with the SMA (MASSES).
The distribution of outflow orientations in the binary pairs is consistent with random or preferentially anti-aligned distributions, demonstrating that these outflows are misaligned.
This result suggests that these systems are possibly formed in environments where the distribution of angular momentum is complex and disordered, and these systems do not come from the same co-rotating structures or from an initial cloud with aligned vectors of angular momentum.

\end{abstract}

\maketitle

\section{Introduction}

Multiplicity is common for both stars \citep[e.g.,][]{2013ARA&A..51..269D} and protostars \citep[e.g.,][]{2013ApJ...768..110C,2000ApJ...529..477L}.
While several scenarios have been proposed to explain the origin of multiplicity, fragmentation at early phases is generally regarded as the main mechanism \citep{2007prpl.conf..133G}.
In particular,
the turbulent fragmentation scenario, which proposes that multiplicity results from turbulent perturbations in a bound core, typically produces wide binaries with separation larger than $\sim$1000 AU \citep[e.g.,][]{2004ApJ...600..769F}.
In contrast, the disk fragmentation scenario, which proposes that fragmentation occurs in gravitationally unstable protostellar disks, produces relatively close binaries with separation typically within a few hundred AU \citep[e.g.,][]{2011ASPC..447...47K}.

A number of theoretical works have studied the alignment between the spin (rotation) axes of binary/multiple (hereafter referred to as multiple for simplicity) components and the binary orbital axis in protostars and stars \citep{2007prpl.conf..395M}.
During the early stages of star formation,
misaligned systems can be produced by turbulent fragmentation where the distribution of angular momentum is complex in the initial core, by dynamical capture in a small cluster, or by ejections in a multiple system \citep{2010ApJ...725.1485O,2012MNRAS.419.3115B}.
At later stages,
misaligned systems can also be produced by effects that alter angular momentum, such as stellar encounters or precession \citep{2000MNRAS.317..773B}.
Alternatively, aligned systems can form from a large co-rotating structure in a massive disk/ring \citep[e.g.,][]{1994MNRAS.269L..45B} or by fragmentation of a core whose angular momentum vectors are aligned.
Aligned systems can also be produced via tidal effects during subsequent evolutionary phases \citep{2000ApJ...538..326L}.
While (mis)alignment at late stages alone cannot provide clear clues in discerning between formation mechanisms, it provides a clearer discriminant at early stages.

\renewcommand\tabcolsep{5pt}
\begin{deluxetable*}{clccccc}
\tablewidth{0pc}
\tablecolumns{7}
\tabletypesize{\scriptsize}
\tablecaption{230 GHz Continuum and Molecular Line Observations}
\tablehead{
    \colhead{No.\tablenotemark{a}} & \colhead{Source\tablenotemark{b}} & \colhead{230 GHz rms\tablenotemark{c}} & \colhead{Molecular\tablenotemark{d}} & \colhead{Chan.\ rms\tablenotemark{d}} & \colhead{Synth.\ Beam\tablenotemark{d}} & \colhead{Map Center\tablenotemark{e}} \\
        \colhead{} & \colhead{} & \colhead{(mJy~beam$^{-1}$)} & \colhead{Line} & \colhead{(mJy~beam$^{-1}$)} & \colhead{($\arcsec$)} & \colhead{($\alpha$, $\delta$ in J2000)}
}
\startdata
1 & Per16 & 2.2 & $^{12}$CO(2-1), (1) & 86 & $4.4\times3.2$ (-15.4$\arcdeg$) & (03:43:51.0, 32:03:16.7)\\
& Per28 & 2.2 & $^{12}$CO(2-1), (1) & 86 & $4.4\times3.2$ (-15.4$\arcdeg$) & (03:43:51.0, 32:03:16.7)\\
\hline
2 & Per26 & 2.7 & $^{12}$CO(2-1), (2) & 139 & $4.1\times3.2$ (-13.9$\arcdeg$) & (03:25:39.0, 30:44:02.0) \\
& Per42 & 2.7 & $^{12}$CO(2-1), (2) & 139 & $4.1\times3.2$ (-13.9$\arcdeg$) & (03:25:39.0, 30:44:02.0) \\
\hline
3 & Per11 & 6.6 & $^{12}$CO(2-1), (3) & 81 & $4.3\times3.2$ (-14.6$\arcdeg$) & (03:43:56.9, 32:03:04.6) \\
\hline
4 & Per33\tablenotemark{f} & 3.4 & $^{12}$CO(2-1), (4) & 60 & $1.3\times0.9$ (86.5$\arcdeg$) & (03:25:36.5, 30:45:22.3) \\
\hline
5 & B1-bN & 2.8 & $^{12}$CO(3-2), (5) & 315 & $3.2\times 2.0$ (-24.6$\arcdeg$) & (03:33:21.0, 31:07:23.8) \\
& B1-bS & 2.8 & $^{12}$CO(3-2), (5) & 315 & $3.2\times 2.0$ (-24.6$\arcdeg$) & (03:33:21.0, 31:07:23.8) \\
& Per41 & 2.8 & $^{12}$CO(3-2), (5) & 315 & $3.2\times 2.0$ (-24.6$\arcdeg$) & (03:33:21.0, 31:07:23.8) \\
\hline
6 & Per8 & 1.9 & $^{12}$CO(2-1), (6) & 104 & $4.1\times 3.8$ (-77.9$\arcdeg$) & (03:44:43.6, 32:01:33.7) \\
& Per55 & 1.9 & $^{12}$CO(2-1), (6) & 104 & $4.1\times 3.8$ (-77.9$\arcdeg$) & (03:44:43.6, 32:01:33.7) \\
\hline
7 & Per12 & 12.7 & $^{12}$CO(3-2), (8) & 197 & $2.8\times2.1$ (-17.2$\arcdeg$) & (03:29:10.5, 31:13:31.0) \\
& Per13 & 11.5 & $^{12}$CO(3-2), (7) & 254 & $3.5\times2.2$ (-23.4$\arcdeg$) & (03:29:12.0, 31:13:01.5) \\
\hline
8 & Per18 & 4.7 & $^{12}$CO(3-2), (10) & 322 & $3.5\times2.1$ (-31.0$\arcdeg$) & (03:29:11.0, 31:18:25.5) \\
& Per21 & 4.7 & $^{12}$CO(3-2), (10) & 322 & $3.5\times2.1$ (-31.0$\arcdeg$) & (03:29:11.0, 31:18:25.5) \\
& Per49 & 2.7 & $^{12}$CO(3-2), (9) & 327 & $3.5\times2.2$ (-31.3$\arcdeg$) & (03:29:12.9, 31:18:14.4) \\
\hline
9 & Per44 & 3.7 & $^{12}$CO(3-2), (11) & 350 & $4.7\times2.2$ (35.9$\arcdeg$) & (03:29:03.4, 31:15:57.7) \\
& SVS 13B & 3.7 & $^{12}$CO(3-2), (11) & 350 & $4.7\times2.2$ (35.9$\arcdeg$) & (03:29:03.4, 31:15:57.7) \\
& SVS 13C & 2.6 & $^{12}$CO(3-2), (12) & 312 & $4.7\times2.2$ (35.8$\arcdeg$) & (03:29:02.0, 31:15:38.1)
\enddata
\label{tbl:data}
\tablenotetext{a}{Wide multiple system number.}
\tablenotetext{b}{PerXX refers to Per-emb-XX in \citet{2009ApJ...692..973E}.}
\tablenotetext{c}{1$\sigma$ noise levels of the 230 GHz continuum observations.}
\tablenotetext{d}{The molecular transitions presented in Fig.~\ref{fig:outflow}, and the 1$\sigma$ noise levels per channel (channel width: 0.5 km~s$^{-1}$) and synthesized FWHM beams (parentheses show position angles measured from North to East) associated with each molecular line observation.  The numbers in parentheses after the molecular transitions indicate the subfigure number where the molecular data is shown in Fig.~\ref{fig:outflow}.}
\tablenotetext{e}{The center of each subfigure in Fig.~\ref{fig:outflow}.}
\tablenotetext{f}{We present the data from the Extended configuration for Per33 published in \citet{2015ApJ...814..114L}.}
\end{deluxetable*}

Therefore, investigating the alignment between the spin axes of multiple components provides important guidance on the formation mechanism of multiple systems.
Among early spectral type, main sequence binaries, most close binaries have aligned spin axes, while wide binaries exhibit misaligned spin axes \citep{2009MNRAS.392..448H}.
In T Tauri disks, a mixture of aligned and misaligned spin axes are observed in wide binaries \citep{2004ApJ...600..789J,2014ApJ...796..120W}.
In the protostellar stage, jet/outflow orientations provide important information for disk orientation since jets are always launched perpendicular to disks while disks are still deeply embedded in envelopes.
However, compared to the studies at later stages, only a few studies have discovered misaligned jets in the youngest objects \citep[e.g.,][]{1993ApJ...408L..49R,1994A&A...285L...1S,2008ApJ...686L.107C}.
To our knowledge, there have been no systematic and statistical studies of the (mis)alignment of protostellar outflows in proto-binary/multiple systems using high-resolution, interferometric observations.

In this Letter we investigate molecular outflows in nine wide multiple systems (projected separation $> 1000$ AU) located in the Perseus Molecular Cloud \citep[distance = 230~pc,][]{2008PASJ...60...37H}.
The data are from a large program with the Submillimeter Array (SMA): Mass Assembly of Stellar Systems and their Evolution \citep[MASSES; PI: Michael Dunham,][]{2015ApJ...814..114L}.
These nine systems cover all the wide systems in the current MASSES sample and cover 70\% of all of the known wide Class 0 multiple systems (some of these systems have Class I components) in Perseus \citep{2016arXiv160100692T}.
With outflows from 23 protostellar objects in these nine systems, they currently provide the largest, unbiased, interferometric sample of outflows in proto-multiple systems observed in the same molecular cloud complex with similar sensitivity, angular resolution, and spectral line coverage.

\section{Observations}

We present data from the Subcompact configuration with the SMA.
The observations were carried out between November 2014 and November 2015.
The observations were obtained in good weather conditions with the zenith opacity at 225 GHz around 0.1.
We observed molecular lines and the continuum at 231.29 GHz and 356.72 GHz simultaneously using the dual receiver mode.
The continuum measurements at the two different frequencies each have an effective bandwidth of 1312 MHz considering the upper and lower sidebands.
High spectral resolution channels were configured for molecular line observations; smoothed velocity resolutions for lines presented in this Letter are the following: 0.5 km~s$^{-1}$ for $^{12}$CO(2-1) (230.53796 GHz) and $^{12}$CO(3-2) (345.79599 GHz), 0.2 km~s$^{-1}$ for C$^{18}$O(2-1) (219.56036 GHz) and N$_{2}$D$^{+}$(3-2) (231.32183 GHz).
We also used the $^{12}$CO(2-1) from the Extended configuration published in \citet{2015ApJ...814..114L} for more clear outflow morphologies in Per33.
The $1\sigma$ rms sensitivities of the 230 GHz continuum and $^{12}$CO observations are summarized in Table~\ref{tbl:data}.

We used the MIR software package\footnote{https://www.cfa.harvard.edu/$\sim$cqi/mircook.html} for data calibration and data reduction.
The uncertainty in the absolute flux calibration was estimated to be $\sim 20$\%.
We used the MIRIAD software package \citep{1995ASPC...77..433S} for data imaging.
The synthesized FWHM beams for the Subcompact data are about $4.3\arcsec \times 3.3\arcsec$ at 230 GHz and $3.0\arcsec \times 2.2\arcsec$ at 345 GHz (see Table~\ref{tbl:data} for details).
These resolutions are sufficient to resolve wide multiples (separation $> 1000$ AU) at the distance of 230~pc to Perseus.
More details about MASSES observations, correlator setup, calibration, and imaging can be found in \citet{2015ApJ...814..114L}.

\section{Results}

\subsection{230 GHz Continuum}
\label{sect:continuum}

\renewcommand\tabcolsep{4pt}
\begin{deluxetable*}{p{0.5in}p{1in}ccr@{$\pm$}lr@{$\pm$}lr@{$\pm$}llr@{$\pm$}lc}
\tabletypesize{\scriptsize}
\tablecaption{Observed Properties of Sources}
\tablewidth{19cm}
\tablecolumns{14}
\tablehead{
    \colhead{Source} & \colhead{Object} & \colhead{R.A.\tablenotemark{a}} & \colhead{Decl.\tablenotemark{a}} & \multicolumn{2}{c}{Peak Int.\tablenotemark{a}} & \multicolumn{2}{c}{$S_{\nu}$\tablenotemark{a}} & \multicolumn{2}{c}{Mass\tablenotemark{b}} & \colhead{Class\tablenotemark{c}} & \multicolumn{2}{c}{Outflow\tablenotemark{d}} & \colhead{V$_{\text{source}}$\tablenotemark{e}} \\
        \colhead{} & \colhead{} & \colhead{(J2000)} & \colhead{(J2000)} & \multicolumn{2}{c}{(mJy~bm$^{\text{-1}}$)} & \multicolumn{2}{c}{(mJy)} & \multicolumn{2}{c}{(M$_{\sun}$)} & \colhead{} & \multicolumn{2}{c}{P.A.(deg)} & \colhead{(km~s$^{\text{-1}}$)}
}
\startdata
Per16 & & 03:43:51.00 & +32:03:23.91 & 21.4&2.3 & 72.5&7.8 & 0.026&0.003 & 0$^{1,2}$ & 7&1\enspace\;(1) & 8.43$\pm$0.02 \\
 Per28 & & 03:43:50.97 & +32:03:08.01 & 16.7&3.0 & 35.8&6.4 & 0.013&0.002 & 0$^{2}$ & 112&2\enspace\;(1) & 8.49$\pm$0.04 \\
 \hline
Per26 & & 03:25:38.87 & +30:44:05.31 & 200.7&11.4 & 310.4&17.6 & 0.113&0.006  & 0$^{1,2}$ & 162&1\enspace\;(1) & 5.14$\pm$0.01 \\
Per42 & & 03:25:39.12 & +30:44:00.45 & 24.2&6.7 & 58.6&16.2 & 0.021&0.006  & I$^{1}$ & 43&2\enspace\;(1) & 5.26$\pm$0.02 \\
\hline
Per11 & IC348 MMS1 & 03:43:57.06 & +32:03:04.66 & 260.8&15.8 & 437.7&26.5 & 0.159&0.01  & 0$^{1,2}$ & 161&1\enspace\;(1) & 8.75$\pm$0.02 \\
     & IC348 MMS2 & 03:43:57.74 & +32:03:10.08 & 43.1&4.6 & 124.4&13.3 & 0.045&0.005  & 0$^{3}$ & 36&12\;(1) & 8.75$\pm$0.03 \\
\hline
Per33 & L1448N-B & 03:25:36.33 & +30:45:14.81 & 423.2&4.6 & 922.4&45.0 & 0.335&0.016  & 0$^{4,5}$ & 122&15\;(1,2) & 5.05$\pm$0.01 \\
    & L1448N-A & 03:25:36.48 & +30:45:21.70 & 66.6&7.6 & 274.4&29.2 & 0.100&0.011  & 0/I$^{4,5}$ & 218&10\;(1,2) & 5.49$\pm$0.01 \\
    & L1448N-NW & 03:25:35.66 & +30:45:34.26 & 68.1&5.5 & 209.4&18.6 & 0.076&0.007  & 0$^{4,5}$ & 128&15\;(1,2) & 3.73$\pm$0.02 \\
\hline
B1-bN & & 03:33:21.20 & +31:07:43.92 & 166.0&6.2 & 209.3&7.8 & 0.076&0.003  & FHSC$^{6}$ & 90&1\enspace\;(3) & 6.92$\pm$0.04 \\
B1-bS & & 03:33:21.34 & +31:07:26.44 & 308.6&12.9 & 353.1&14.8 & 0.128&0.005  & FHSC$^{6}$ & 112&6\enspace\;(3) & 6.45$\pm$0.07 \\
Per41\tablenotemark{f} & & 03:33:20.34 & +31:07:21.36 &  \nodata & \nodata & \nodata & \nodata  & \nodata & \nodata & I$^{1}$ & 30&5\enspace\;(1) & \nodata\tablenotemark{h} \\
\hline
Per8 & & 03:44:43.98 & +32:01:34.97 & 125.9&7.2 & 159.8&9.1 & 0.058&0.003  & 0$^{1}$ & 15&5\enspace\;(4) & 10.46$\pm$0.04 \\
Per55\tablenotemark{f} & & 03:44:43.30 & +32:01:31.24 & \nodata & \nodata & \nodata & \nodata  & \nodata & \nodata & I$^{1}$ & 115&2\enspace\;(1) & 11.18$\pm$0.05 \\
\hline
Per12 & & 03:29:10.50 & +31:13:31.33 & 2722.0&82.6 & 4200.0&127.5 & 1.524&0.046 & 0$^{1,2}$  & 19&5\tablenotemark{h}\;(5,6) & 6.70$\pm$0.01 \\
Per13 & NGC1333 IRAS4B & 03:29:11.99 & +31:13:08.14 & 915.0&32.5 & 1202.0&42.7 & 0.436&0.015  & 0$^{1,2}$ & 176&2\enspace\;(1) & 6.76$\pm$0.01 \\
      & NGC1333 IRAS4B' & 03:29:12.82 & +31:13:07.00 & 330.1&30.0 & 419.5&38.1 & 0.152&0.014  & 0$^{1,2}$ & 90&1\enspace\;(1,7) & 6.99$\pm$0.04 \\
\hline
Per18 & & 03:29:11.26 & +31:18:31.33 & 130.2&7.7 & 178.4&10.6 & 0.065&0.004  & 0$^{1,2}$ & 150&1\enspace\;(1) & 8.09$\pm$0.02 \\
Per21 & & 03:29:10.69 & +31:18:20.11 & 71.3&6.2 & 154.6&13.4 & 0.056&0.005  & 0$^{1,2}$ & 48&13\;(1) & 8.79$\pm$0.04 \\
Per49 & & 03:29:12.90 & +31:18:13.87 & 14.8&2.4 & 23.2&3.8 & 0.008&0.001  & I$^{1}$ & 27&2\enspace\;(1) & \nodata\tablenotemark{i} \\
\hline
Per44 & SVS 13A1\tablenotemark{g} & 03:29:03.75 & +31:16:03.59 & 424.6&19.8 & 525.7&24.5 & 0.191&0.009  & 0/I$^{7}$ & 130&5\enspace\;(8) & \nodata\tablenotemark{j} \\
    & SVS 13A2\tablenotemark{g} & 03:29:03.40 & +31:16:00.10 & 80.06&13.3 & 232.7&38.7 & 0.084&0.014  & 0/I$^{7}$ & \nodata & \nodata & \nodata\tablenotemark{j} \\
    SVS 13B & & 03:29:03.04 & +31:15:51.47 & 332.5&20.8 & 663.9&41.5 & 0.241&0.015  & 0$^{7}$ & 170&10\;(1,9) & \nodata\tablenotemark{j} \\
    SVS 13C & & 03:29:02.00 & +31:15:38.31 & 69.9&7.8 & 139.7&15.6 & 0.051&0.006  & 0$^{7}$ & 0&1\enspace\;(1) & \nodata\tablenotemark{j}
\enddata
\label{tbl:continuum}
\tablenotetext{a}{R.A., Decl., peak intensity, and total flux density ($S_{\nu}$) were obtained by fitting Gaussians to the 230 GHz continuum primary-beam corrected maps using MIRIAD task \textit{imfit}.}
\tablenotetext{b}{The evolutionary class of each object. $^{1}$\citet{2009ApJ...692..973E}. $^{2}$\citet{2014ApJ...787L..18S}. $^{3}$\citet{2014MNRAS.444..833P}. $^{4}$\citet{1998ApJ...509..733B}. $^{5}$\citet{2006AJ....131.2601O}. $^{6}$FHSC: First Hydrostatic Core Candidate; \citet{2012A&A...547A..54P}. $^{7}$\citet{2009ApJ...691.1729C}.  NGC1333 IRAS4B' is assumed to have the same class as IRAS4B, and SVS 13A2 is assumed to have the same class as SVS 13A1.}
\tablenotetext{c}{Envelope mass considering dust and gas.}
\tablenotetext{d}{Position angles of outflows (measured from north to east).  Parentheses show the references for outflow identification: (1) This work. (2) \citet{2015ApJ...814..114L}. (3) \citet{2015A&A...577L...2G}. (4) \L.\ Tychoniec et al., (in prep) (5) \citet{2005ApJ...630..976C}. (6) \citet{2015A&A...584A.126S}. (7) \citet{2014MNRAS.444..833P}. (8) \citet{2000A&A...362L..33B}. (9) \citet{1998A&A...339L..49B}. The angles and uncertainties were determined based on manual identification from multiple people.}
\tablenotetext{e}{Velocities of sources obtained from fitting of C$^{18}$O(2-1) or N$_{2}$D$^{+}$(3-2) spectra averaged over one synthesized beam at continuum peaks.  B1-bN and B1-bS used N$_{2}$D$^{+}$, and the rest used C$^{18}$O.}
\tablenotetext{f}{Per41 and Per55 are not detected at 230 GHz continuum.  The coordinates reported here are from \citet{2009ApJ...692..973E}.}
\tablenotetext{g}{SVS 13A1 and SVS 13A2 are not resolved in the continuum map and appear as one extended structure.  We fit two Gaussian profiles to obtain the reported coordinates.}
\tablenotetext{h}{Per12 (NGC1333 IRAS4A) contains two sources, 4A1 and 4A2, and each drives an outflow \citep{2005ApJ...630..976C}.  We use the outflow from 4A2.}
\tablenotetext{i}{The velocity could not be obtained due to poor signal-to-noise ratios in the spectrum.}
\tablenotetext{j}{C$^{18}$O observations were not performed due to correlator issues.}
\end{deluxetable*}

Table~\ref{tbl:continuum} lists the properties of the observed sources derived from the 230 GHz continuum.
We observed 19 protostellar sources (listed as ``Source" in Table~\ref{tbl:continuum}) including 15 sources identified by a 1.1~mm Bolocam continuum and \textit{Spitzer} survey \citep{2009ApJ...692..973E}, SVS 13B and SVS 13C \citep{2000ApJ...529..477L}, and two first hydrostatic core candidates B1-bN and B1-bS \citep{2012A&A...547A..54P}.
Some of these sources contain multiple sources as revealed by higher angular resolution observations, and we list those multiples as ``Object" in Table~\ref{tbl:continuum}.
In total, there are 24 protostellar objects.
Our 230 GHz continuum observations detected 22 of these 24 protostellar objects (except for Per41 and Per55).

These objects form nine wide multiple systems with separations larger than 1000 AU.
Several objects (Per18, L1448N-B, L1448N-NW, SVS 13A1, Per12, IC348 MMS) contain close binaries with separations less than a few hundred AU \citep{2016arXiv160100692T}.
In this Letter we focus on the properties of the wide systems and regard each close system as one source with properties based on our 230 GHz continuum observations.

The position, peak intensity, and total flux density of each object were obtained by fitting a Gaussian to the 230 GHz continuum images.
We derived the masses of the envelopes using the equation (assuming the 230 GHz dust emission is optically thin):
$$\text{Mass} = \dfrac{S_{\nu} d^{2}}{\kappa_{\nu} B_{\nu}(T)}, $$ where $S_{\nu}$ is the total flux density at 230 GHz, $d$ is the distance to the object (230~pc), $\kappa_{\nu}$ is the dust opacity derived from $\kappa_{\nu} = 0.1 \times (\nu / 10^{3} \ \text{GHz})^{\beta}$ cm$^{2}$~g$^{-1}$ \citep{1990AJ.....99..924B} with an assumed gas-to-dust ratio of 100, and $B_{\nu}(T)$ is the blackbody intensity at dust temperature T.
We assumed a dust temperature of 30 K and $\beta = 1.2$ \citep[e.g.,][]{2007ApJ...659..479J,2001ApJ...546L..49S}.
The masses range from 0.01 M$_{\sun}$ to 1.5 M$_{\sun}$ (Table~\ref{tbl:continuum}).
The uncertainties of these mass estimates are at least a factor of 2 due to the choices of dust opacity, dust temperature, the gas-to-dust ratio, the beta value, and resolved-out contributions from emission at larger scales \citep[e.g.,][]{2014MNRAS.444..887D}.

\subsection{Outflow Identification}
\begin{figure*}
\includegraphics[scale=0.76]{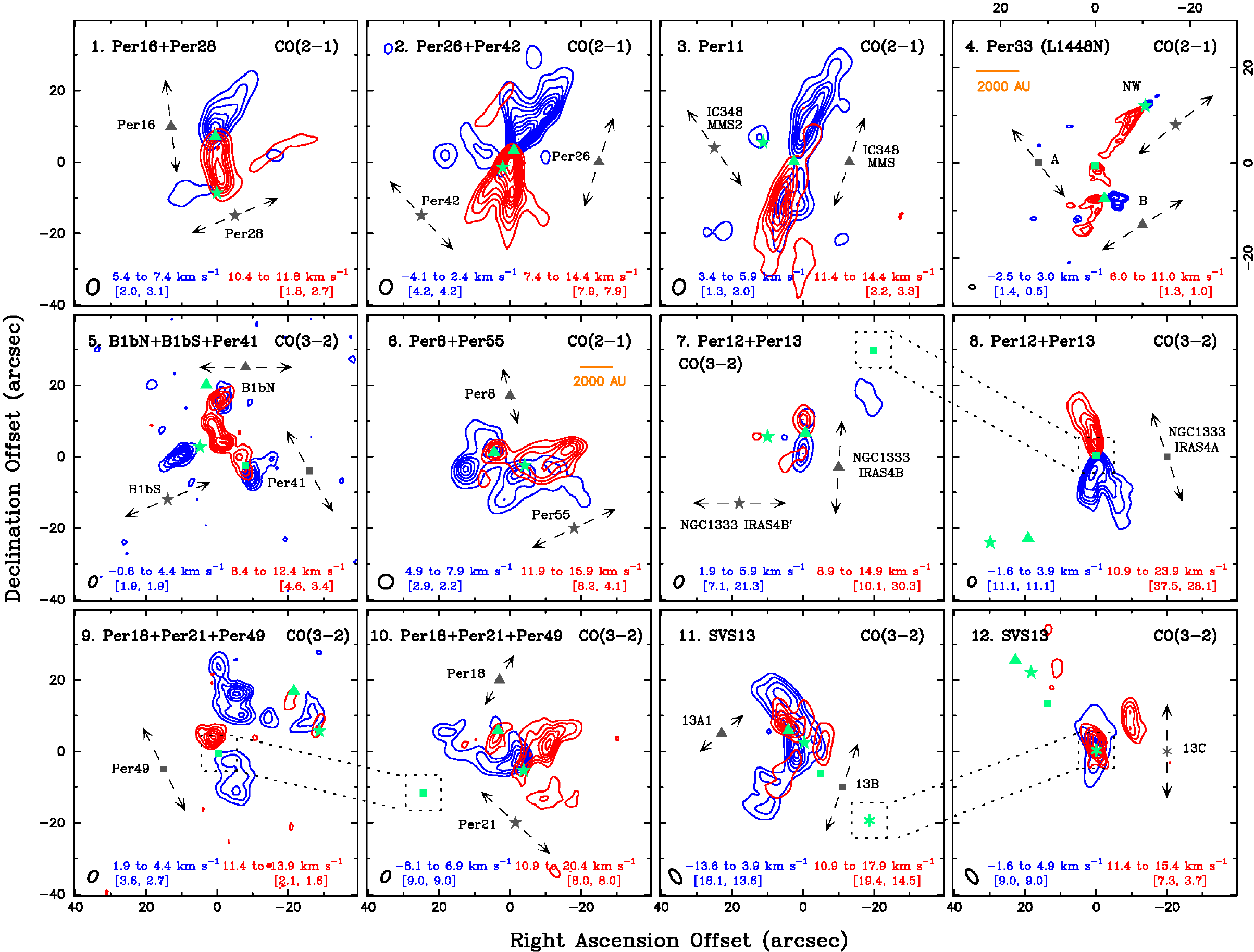}
\caption{
\scriptsize
Integrated intensity maps of the redshifted and blueshifted outflows from the $^{12}$CO(2-1) or $^{12}$CO(3-2) data (indicated at upper left corners) in the nine multiple systems.
The data are from the Subcompact configuration except Per33 (top right corner), which shows the Extended data.
The numbers at the upper-left corner in each map indicates the subfigure number.
The offset coordinates are relative to the phase centers reported in Table~\ref{tbl:data}.
The green symbols are the positions of the 230 GHz continuum peaks from Table~\ref{tbl:continuum}.
The grey symbols and arrows show the orientation of the outflow from each protostar indicated by the corresponding green symbols.
The dashed lines are the identified outflow orientations (Table~\ref{tbl:continuum}).
The integrated velocity ranges are shown near the bottom in red (blue) corresponding to the red (blue) lobes.
The values of the first contour and the subsequent contour steps for the red (blue) lobe of each outflow are given inside brackets in units of Jy~beam$^{-1}$~km~s$^{-1}$.
The synthesized beam is drawn at the bottom-left corner of each panel.
Each panel has the same box size except for Per33, and the solid, orange lines indicate a scale of 2000 AU.
}
\label{fig:outflow}
\end{figure*}

Figure~\ref{fig:outflow} shows outflows from the nine wide multiple systems.
We inspected both $^{12}$CO(2-1) and $^{12}$CO(3-2) maps for each object and show the molecular line transition that presents the clearest outflow morphologies.
We identified outflows primarily based on our $^{12}$CO data; we also investigated \textit{Spitzer-IRAC} 4.5~$\micron$ images \citep{2006ApJ...645.1246J} to examine scattered light from outflow cavities and confirm the identifications.
The majority of the objects have clear blue- and red-shifted emissions offset from the protostar in position (hereafter ``blue-lobe" and ``red-lobe") including Per16, Per28, Per26, Per42, IC348 MMS, L1448N-B, L1448N-A, L1448N-NW, Per41, NGC1333 IRAS4B, and Per49.
IC348 MMS2 shows detection in the blue lobe toward the north-east, a feature also observed in \citet{2014MNRAS.444..833P} with a consistent position and velocity range.
NGC1333 IRAS4B' shows a weak outflow in the E-W direction, consistent with the detection in \citet{2014ApJS..213...13H}.
The outflow identification of Per18 is based on the strong red lobe as a jet-like morphology is also observed in the \textit{IRAC} 4.5~$\micron$ image with an orientation consistent with the red lobe.
The blue lobe of Per21 exhibits an arc shape, while the red lobe is less clear and is not symmetric about the source.
The \textit{IRAC} 4.5~$\micron$ image of Per21 shows clear outflow structures that agree with the identified orientation based on $^{12}$CO.
SVS 13B shows a red lobe approximately in the N-S direction, and \citet{1998A&A...339L..49B} observed a blue lobe in the velocity range from -17 to 6.5 km~s$^{-1}$ with a consistent orientation.
SVS 13C has overlapping blue and red lobes along the line of sight, suggesting that the outflow is pole-on.

We used identifications from the literature for a few objects where we do not observe clear outflow morphologies in the $^{12}$CO maps.
B1-bN and B1-bS exhibit complicated CO outflows, particularly in the redshifted emissions between B1-bN and B1-bS \citep{2014ApJ...789...50H}.
We used the H$_{2}$CO and CH$_{3}$OH observations in \citet{2015A&A...577L...2G}, which showed two clear outflows in approximately E-W directions.
Per12 (NGC1333 IRAS4A) contains a close binary, 4A1 and 4A2, and each source drives an outflow \citep{2005ApJ...630..976C}.
We used the position angle of the outflow from A2 since it is stronger than the outflow from A1 \citep{2015A&A...584A.126S,2016arXiv160105229C}.
For SVS 13A1, we used the identification from \citet{2000A&A...362L..33B}, which was based on the $^{12}$CO(2-1) outflow sensitive to extremely high velocities and a chain of Herbig-Haro objects in HH 7-11.
Table~\ref{tbl:continuum} lists the position angles of the identified outflows associated with each object.

We used C$^{18}$O(2-1) and N$_{2}$D$^{+}$(3-2) data to obtain the velocity of each object.
Our data show that N$_{2}$D$^{+}$ peaks coincide well with continuum peaks in FHSCs, while C$^{18}$O peaks coincide well with continuum peaks in Class 0 and I objects.
This is expected from chemistry since CO is evaporated to detect and destroys N$_{2}$H$^{+}$/N$_{2}$D$^{+}$ in more evolved sources, while insufficient CO is evaporated to detect for FHSCs.
Therefore, we used N$_{2}$D$^{+}$ to obtain the source velocity for B1-bN and B1-bS and used C$^{18}$O for the rest of the objects.
Each source velocity was obtained by fitting a Gaussian profile to the spectrum averaged over one synthesized beam at the continuum peak.
The source velocities are listed in Table~\ref{tbl:continuum}.

\section{Implications}

\subsection{Gravitational Boundedness}

\renewcommand\tabcolsep{3pt}
\begin{deluxetable}{lcr@{$\pm$}lr@{$\pm$}lr@{$\pm$}l}
\tablewidth{0pc}
\tablecolumns{8}
\tabletypesize{\scriptsize}
\tablecaption{Physical Properties of Each Pair}
\tablehead{
\colhead{Pair} & \colhead{$\Delta d$} & \multicolumn{2}{c}{V$_{e}$\tablenotemark{a}} & \multicolumn{2}{c}{$\Delta V$\tablenotemark{b}} & \multicolumn{2}{c}{$\Delta \theta$\tablenotemark{c}} \\
\colhead{} & \colhead{(AU)} & \multicolumn{2}{c}{(km~s$^{\text{-1}}$)} & \multicolumn{2}{c}{(km~s$^{\text{-1}}$)} & \multicolumn{2}{c}{(deg)}
}
\startdata
Per16+Per28 & 3658 & 0.14&0.006 & 0.06&0.04 & 75&2 \\
Per26+Per42 & 1863 & 0.36&0.011 & 0.12&0.02 & 61&2 \\
IC348 MMS+MMS2 & 2347 & 0.39&0.011 & $<$0.01&0.04 & 55&12 \\
L1448N-B+N-A & 1646 & 0.68&0.015 & 0.43&0.01 & 84&18 \\
L1448N-B+N-NW & 4895 & 0.38&0.008 & 1.32&0.02 & 6&21 \\
L1448N-A+N-NW & 3776 & 0.29&0.011 & 1.75&0.02 & 90&18 \\
B1-bN+B1-bS & 4042 & 0.30&0.004 & 0.47&0.08 & 22&6 \\
B1-bN+Per41 & 5777 & \nodata&\nodata & \nodata&\nodata & 60&5 \\
B1-bS+Per41 & 3176 & \nodata&\nodata & \nodata&\nodata & 82&8 \\
Per8+Per55 & 2166 & \nodata&\nodata & 0.72&0.06 & 80&5 \\
NGC1333 IRAS4A+4B & 6912 & 0.71&0.009 & 0.06&0.01 & 23&5 \\
NGC1333 IRAS4A+4B' & 8841 & 0.58&0.008 & 0.29&0.04 & 61&5 \\
NGC1333 IRAS4B+4B' & 2463 & 0.65&0.011 & 0.24&0.04 & 86&2 \\
Per18+Per21 & 3079 & 0.26&0.007 & 0.70&0.04 & 78&13 \\
Per18+Per49 & 6285 & 0.14&0.004 & \nodata&\nodata & 78&13 \\
Per21+Per49 & 6671 & 0.13&0.005 & \nodata&\nodata & 21&13 \\
SVS 13A1+13B & 3486 & 0.47&0.009 & \nodata&\nodata & 40&11 \\
SVS 13A1+13C & 7774 & 0.23&0.005 & \nodata&\nodata & 50&5 \\
SVS 13B+13C & 4309 & 0.35&0.010 & \nodata&\nodata & 10&10
\enddata
\tablenotetext{a}{Escape velocity in this two-body system.}
\tablenotetext{b}{Difference in the velocities of the two sources (see Table~\ref{tbl:continuum}).}
\tablenotetext{c}{Difference between the position angles of two outflows (see Table~\ref{tbl:continuum}).}
\label{tbl:pair}
\end{deluxetable}

Dynamical interactions play a crucial role in the alignment of multiple systems \citep{2007prpl.conf..395M}.
To better understand the dynamics in our sample,
we investigated gravitational boundedness in each system.
We calculated the escape velocity for each binary pair in a two-body system: $V_{e} = \sqrt{2G(M_{1}+M_{2})/\Delta d}$, where $G$ is the gravitational constant, $M_{1}$ and $M_{2}$ are the masses of the two objects in a pair, and $\Delta d$ is the separation between the two objects.
We used the masses calculated in Table~\ref{tbl:continuum} for $M_{1}$ and $M_{2}$.
The projected separations and the resulting escape velocity are listed in Table~\ref{tbl:pair}.

The velocity difference ($\Delta V$) between two objects in a pair is calculated based on the source velocities in Table~\ref{tbl:continuum}; the results are listed in Table~\ref{tbl:pair}.
These results show that the majority of the pairs have $V_{e}$ larger than $\Delta V$ by a factor of $2-3$, implying that most systems are bound.
However, this comparison is highly uncertain due to several factors.
First, the separations we used in the formula are the projected separations, which likely under-estimate the actual separations.
With actual separations being larger, the systems would be more unbound.
In addition, the derived $\Delta V$ only considers components along the line of sight; components in other directions would increase the differences in the velocities, and the systems would be more unbound.
Furthermore, the envelope mass estimates are uncertain by a factor of at least a few due to the choices of dust temperature and dust opacity (Sect.~\ref{sect:continuum}), and the masses from the central protostars are not included in the envelope masses.
By including the masses from protostars the total masses could increase by a factor of two or more.
The mass estimates also suffer from spatial filtering and do not have contributions from large scales.
We argue that the level of contributions the interferometric observations resolve out are similar in both $M_{1}$ and $M_{2}$, and our $V_{e}$ estimates based on these masses would be lower limits considering this effect.
An increase in the masses would cause the systems to be more bound.
With these uncertainties, the estimates of escape velocities can easily be altered by a factor of a few.

Considering these uncertainties, we speculate that the systems with $V_{e} > \Delta V$ may be soft binaries (loosely bound) or intermediate binaries (between loosely bound and tightly bound; \citet{2014prpl.conf..267R}).
Soft binaries are likely to be destroyed by an encounter \citep{2011MNRAS.415.1179M}.
Intermediate binaries may be disrupted or may survive depending on the details of the individual dynamical history \citep{2012MNRAS.424..272P}.

\subsection{Outflow Orientation}

\begin{figure}
\includegraphics[scale=0.35]{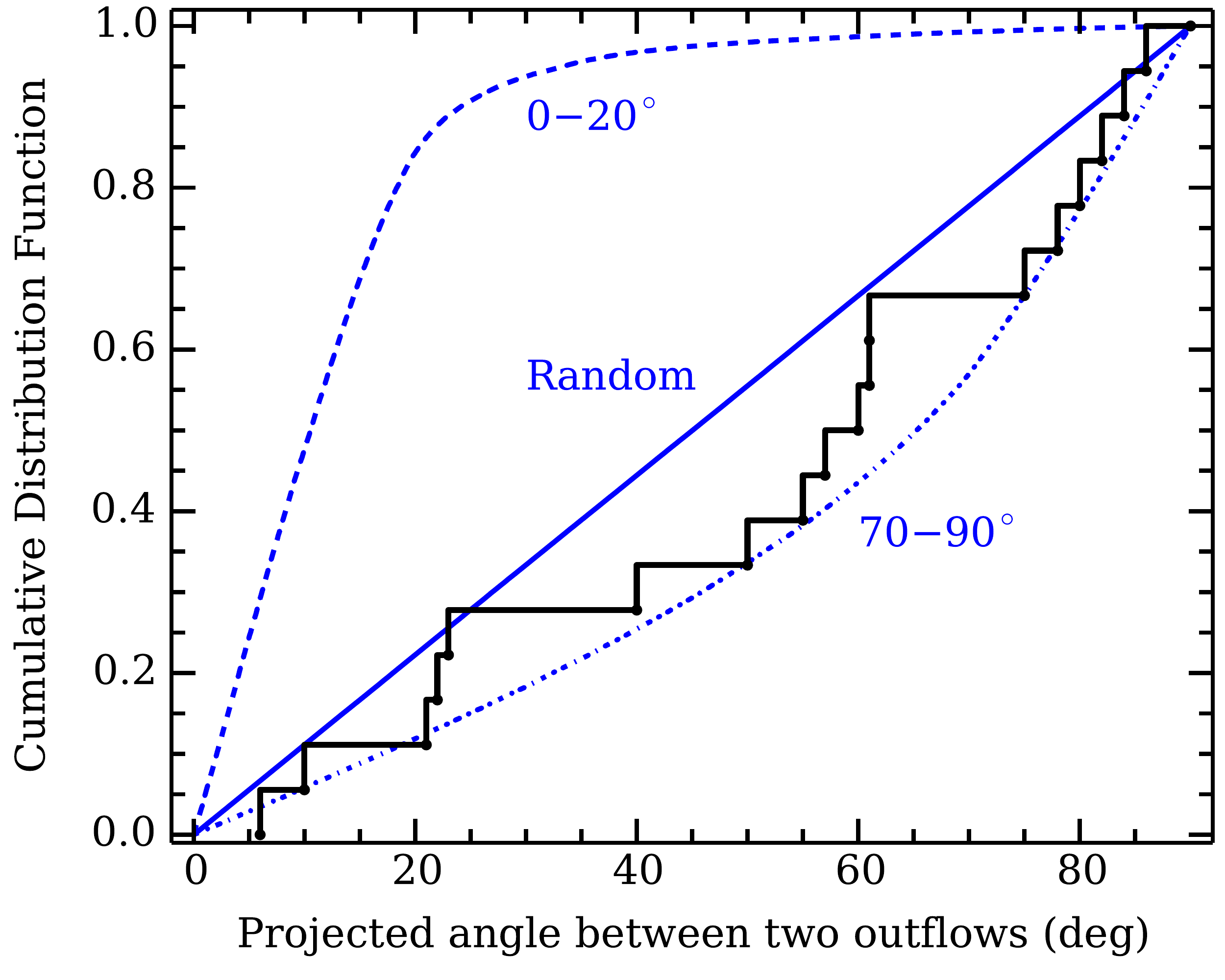}
\caption{
\scriptsize
The cumulative distribution functions of projected outflow orientation differences from the tightly aligned distribution (dashed blue line), random distribution (solid blue line), and preferentially anti-aligned distribution (dash-dot blue line) generated by 3D Monte Carlo Simulations.  The black line is the outflow orientation differences from all the pairs in Table~\ref{tbl:pair}.
}
\label{fig:ks}
\end{figure}

When comparing the differences in outflow orientations, we consider all the possible pairs in each system with separations larger than 1000 AU and less than 10000 AU \citep{2016arXiv160100692T}.
Table~\ref{tbl:pair} lists the difference in the outflow orientation for each pair ($\Delta \theta$) derived from the data in Table~\ref{tbl:continuum}.
To investigate if the distribution of observed outflow orientation differences, which are projected on the plane of the sky, reflects particular intrinsic distributions in the 3D space, we performed Monte Carlo simulations considering three distributions in 3D: tightly aligned (outflows orientation differences less than 20$\arcdeg$), random, and preferentially anti-aligned (outflow orientation differences between 70$\arcdeg$ and 90$\arcdeg$).
We then project these outflows generated in 3D onto the plane of the sky in 2D.

Figure~\ref{fig:ks} shows the cumulative distribution functions of the projected outflow orientation differences from the three distributions in 3D.
The black solid line shows the observed data of outflow orientation differences from Table~\ref{tbl:pair}.
We performed Kolmogorov-Smirnov (K-S) tests with three null hypotheses: the observed distribution is the same as the tightly aligned, random, and preferentially anti-aligned distributions, respectively.
The p-values from the K-S tests are $1.5\times10^{-10}$ for the tightly aligned distribution, 0.18 for the random distribution, and 0.5 for the anti-aligned distribution.
By adopting a significance level of 0.05,
we reject the null hypothesis that the observed distribution is the same as the tightly aligned distribution.
However, we cannot reject the other two null hypotheses.
This result suggests that the outflows in these multiple systems are misaligned.

Provided that most of our objects are at the youngest Class 0 stage, our observations are likely the best available probe of the initial conditions of wide multiple formation.
Our K-S test results suggest that members of these wide multiple systems do not come from the same co-rotating structures, or from an initial cloud with aligned vectors of angular momentum.
The results suggest that these wide multiple systems likely formed in environments where the distribution of angular momentum was complex and disordered.
One major possibility for such an environment is turbulent fragmentation, where the distribution of angular momentum has spatial variations \citep{2000MNRAS.317..773B}.
In this case misaligned outflows are expected in wide systems \citep{2012MNRAS.419.3115B,2011ApJ...743...91O}.
Another possibility is dynamical interactions such as dissipative star-disc encounters via capture of a passing object, typically with a different direction of angular momentum \citep[e.g.,][]{1991MNRAS.249..584C}.
However, this is less likely since the frequency for such favorable encounters is low \citep{2002ARA&A..40..349T}.

\section{Acknowledgment}

This work is based primarily on observations obtained with the SMA, a joint project between the Smithsonian Astrophysical Observatory and the Academia Sinica Institute of Astronomy and Astrophysics and funded by the Smithsonian Institution and the Academia Sinica.
The authors thank the SMA staff for executing these observations as part of the queue schedule, and Charlie Qi and Mark Gurwell for their technical assistance with the SMA data.

K.I.L.\ acknowledges support from NASA grant NNX14AG96G.
M.M.D.\ acknowledges support from NASA ADAP grant NNX13AE54G and from the Submillimeter Array through an SMA postdoctoral fellowship.
T.L.B.\ also acknowledges partial support from NASA ADAP grant NNX13AE54G.
E.I.V.\ acknowledges support from the Russian Ministry of Education and Science grant 3.961.2014/K.
S.S.R.O.\ acknowledges support from the National Aeronautics and Space Administration under Grant No.\ 14-ATP14-0078 issued through the Astrophysics Theory Program.
J.J.T.\ is currently supported by grant 639.041.439 from the Netherlands Organisation for Scientific Research (NWO).

\bibliographystyle{apj}

\end{document}